# Current driven asymmetric magnetization switching in perpendicularly magnetized CoFeB/MgO heterostructures

Jacob Torrejon[1,2], Felipe Garcia-Sanchez[3], Tomohiro Taniguchi[4], Jaivardhan Sinha[1], Seiji Mitani[1], Joo-Von Kim[3], and Masamitsu Hayashi[1*]

[1]*National Institute for Materials Science, Tsukuba 305-0047, Japan*
[2]*Unité Mixte de Physique CNRS/Thales, 1 Avenue Augustin Fresnel, 91767 Palaiseau, France*
[3]*Institut d'Electronique Fondamentale, UMR CNRS 8622, Université Paris-Sud, 91405 Orsay, France*
[4]*National Institute of Advanced Industrial Science and Technology (AIST), Spintronics Research Center, Tsukuba, Ibaraki 305-8568, Japan*

The flow of in-plane current through ultrathin magnetic heterostructures can cause magnetization switching or domain wall nucleation owing to bulk and interfacial effects. Within the magnetic layer, the current can create magnetic instabilities via spin transfer torques (STT). At interface(s), spin current generated from the spin Hall effect in a neighboring layer can exert torques, referred to as the spin Hall torques, on the magnetic moments. Here, we study current induced magnetization switching in perpendicularly magnetized CoFeB/MgO heterostructures with a heavy metal (HM) underlayer. Depending on the thickness of the HM underlayer, we find distinct differences in the in-plane field dependence of the threshold switching current. The STT is likely responsible for the magnetization reversal for the thinner underlayer films whereas the spin Hall torques cause the switching for thicker underlayer films. For the latter, we find differences in the switching current for positive and negative currents and initial magnetization directions. We find that the growth process during the film deposition introduces an anisotropy that breaks the symmetry of the system and causes the asymmetric switching. The presence of such symmetry breaking anisotropy enables deterministic magnetization switching at zero external fields.

*Email: hayashi.masamitsu@nims.go.jp

# I. Introduction

Spin transfer torques (STT), which represent the transfer of spin angular momentum from a spin-polarized current to local magnetization, are now well-established for their use to control magnetization [1, 2]. STT has been exploited in magnetic tunnel junctions (MTJs) for developing advanced non-volatile memory (MRAM). One of the main challenges to achieve reliable operation of MRAM is to increase the margin of reading and writing current, which requires high magnetoresistance ratio and low writing current.

Alternatively, a three terminal device can be used to overcome this problem by separating the circuit for reading and writing [3-7]. For such device, one can make use of the recently discovered spin orbit effects to trigger magnetization switching [3, 8]. In particular, the spin Hall effect (SHE) in heavy metal (HM) layers [9] can generate sufficiently large spin current to manipulate magnetic moments of a magnetic layer adjacent to the HM layer. The torque on the magnetic moments exerted by the spin current is referred to as the spin Hall torque. Intuitively, the action of STT and spin Hall torques on magnetization is governed by the same physics, however, the underlying processes related to the latter and the difference between the two torques are not clear and require further thorough study [10-14].

For STT driven magnetization switching, it is beneficial to use MTJs with perpendicularly magnetized "free" layer to achieve fast and low current magnetization switching [15-17]. With regard to magnetization switching of a perpendicularly magnetized layer via the spin Hall torque, one needs to apply an in-plane field directed along the current in order to reverse the magnetization direction [3, 8, 18]. The need to apply such in-plane field may require additional costly processing for developing devices and thus would preferably be avoided. On this front, it has been recently demonstrated that magnetization switching can be triggered via the spin Hall torque in the absence of any magnetic field by using sophisticated device structuring[19, 20]. In order to fully utilize spin Hall torque driven magnetization switching for technological device applications, the underlying physics of the switching process needs to be further clarified.

Here we report magnetization switching in wires patterned from CoFeB|MgO heterostructures with heavy metal (HM) underlayers. We study the threshold current needed to reverse magnetization as a function of pulse amplitude, pulse length and in-plane magnetic field. Distinct differences are found in the in-plane field dependence of the switching current between STT- and spin Hall torque-driven processes. Direct current flowing through the magnetic layer can cause instability of the magnetic moments via STT and consequently can result in magnetization switching, however, with no difference in the switching probability against the current flow direction or initial magnetization direction. In contrast, for spin Hall torque driven magnetization switching, the switching current is different for positive and negative currents and initial magnetization directions. We find that a tilt in the uniaxial anisotropy axis, first reported by You *et al.*[20] to show that such effect enables spin Hall torque switching at zero field, develops during the film deposition process and is found to be responsible for the asymmetric magnetization switching with current.

## II. Experimental results

### A. Experimental setup

The heterostructures studied here are the same with those reported in Ref. [21]. The film stack Sub.|$d$ HM|1 CoFeB|2 MgO|1 Ta (figures indicate film thicknesses in nanometers) is sputtered onto thermally oxidized Si substrates (SiO$_2$ is 100 nm thick). We have studied a number of materials for the HM underlayer (TaN, Hf, W) and found similar results. Representative results from the TaN underlayer films are mostly reported here. TaN is formed by reactively sputtering Ta in a mixed gas atmosphere of Ar and N$_2$[22]: the atomic concentration is Ta$_{48\pm5}$N$_{52\pm5}$ for the results shown here. The underlayer thickness $d$ is varied within the substrate using a linear shutter during the sputtering. Wires are patterned using optical lithography and Ar ion etching and a subsequent lift-off process is employed to form electrical contacts made by 10 Ta|100 Au (units in nanometers). The width and the length of the patterned wires are 5 μm and 20-30 μm, respectively.

Figure 1(a) shows a typical optical microscopy image of the patterned wires and the definition of the coordinate axes. A pulse generator is connected to one of the contacts to apply constant amplitude voltage pulses (0.5-100 ns long) to the wire. Positive current corresponds to current flow along the +$x$ direction. We use Kerr microscopy to study magnetization reversal driven by magnetic field and/or current.

The magnetic easy axis of the films points along the film normal owing to the perpendicular magnetic anisotropy (PMA) developed at the CoFeB|MgO interface[15, 22]. Figure 1(b) shows magnetization hysteresis loops of two TaN underlayer films measured using Kerr microscopy. $H_{SW}$, the average (absolute) out of plane field ($H_Z$) needed to switch the magnetization from +$z$ to −$z$ and vice versa, is ~100 Oe for the two films shown in Fig. 1(b): typical values of $H_{SW}$ range between ~50-500 Oe for all films studied. Note that $H_{SW}$ represents the field needed to nucleate reversed domains; once a reversed domain forms, domain wall propagation takes place to magnetize the entire wire (the wall propagation field is ~5 to ~30 Oe). The variation of $H_{SW}$ is mostly related to the strength of PMA for each film: the magnetic and electrical properties of all films studied here can be found in Ref [21].

Current-induced magnetization switching is studied using the captured Kerr images. To determine the threshold current for magnetization switching, the following sequence is performed. (1) A large out of plane field ($H_Z$) is applied to uniformly magnetize the wire along the $z$ direction. (2) The out of plane field is reduced, typically to zero unless noted otherwise, and an in-plane field directed along $x$ ($H_X$) or $y$ ($H_Y$) is applied. Then a Kerr image of the uniform state is captured to obtain a reference image. (3) Current is injected to the wire by applying voltage pulse(s) from the pulse generator. The pulse is either a single pulse or a pulse train with each pulse separated in time by ~10 ms. The pulse length is fixed to 100 ns unless noted otherwise (Fig. 9a). After the application of the voltage pulse(s), a second Kerr image is captured. The first image captured in (2) is subtracted from this second image to acquire the "subtracted image", which we use to calculate the area where the magnetization direction reversed upon the pulse application. The switching probability ($P_{SW}$) is calculated by dividing the area where the magnetization switched with the area of the wire. This process (1-3) is repeated 5 times to acquire statistics: the switching probability shown hereafter corresponds to the mean of $P_{SW}$ of the 5 measurements.

## B. Current-induced magnetization reversal

Figures 1(c) and 1(d) display the probability of magnetization switching as a function of pulse amplitude for the two devices shown in Fig. 1(b). For illustration purposes, we multiply the probability by –1 when the initial magnetization direction points along the –z direction (red circles). At the corners of each graph, representative Kerr images corresponding to the maximum pulse amplitude for both current directions and initial magnetization configurations are shown. From these images, it can be seen that the switching characteristics depend on the film structure. Wires with thin TaN underlayers (Fig. 1(c)) show a symmetric nucleation process with respect to the current flow direction and the initial magnetization configuration: above the threshold voltage, the switching probability increases and saturates at ~0.5. For thicker TaN underlayer films (Fig. 1(d)) the probability is asymmetric with respect to the current direction and the initial magnetization configuration. For initial magnetic states pointing along +z (–z), the switching probability is lower for negative (positive) current.

These results indicate that different mechanisms are involved in the magnetization reversal process depending on the thickness of the underlayer. Figure 2 shows the TaN underlayer thickness dependence of the threshold current density ($J_N^C$) that flows through the underlayer. We define $J_N^C$ as the minimum current density needed to achieve switching probability exceeding 25%. $J_N^C$ is calculated using the threshold pulse amplitude, the resistance of the wire, the thickness and the resistivity ($\rho$) of the CoFeB layer ($\rho$~160 $\mu\Omega$ cm) and the HM underlayer ($\rho$~375 $\mu\Omega$ cm for $Ta_{48}N_{52}$)[21]. The solid and open symbols in Figs. 2 represent positive and negative $J_N^C$, respectively; here we show $-J_N^C$ for negative current to compare the absolute value with that of positive current. The dependence of $J_N^C$ on the initial magnetization states is shown in Figs. 2(a) and 2(b). The asymmetry in the threshold current density with respect to the current flow direction and the initial magnetization direction reduces to near zero when the TaN underlayer thickness is below ~1 nm. The degree of asymmetry is nearly constant when the underlayer thickness is larger than ~2 nm. This trend qualitatively agrees with the underlayer thickness dependence of the "effective field" due to the spin Hall torque[21, 23] (see Figs. 4(a) and 4(b)). When the thickness of the TaN underlayer is thinner than its spin diffusion length, the effective field is nearly zero. In contrast, if the underlayer thickness is larger than its spin diffusion length, ~2.5 nm for TaN[13, 21], the effective field saturates and becomes constant against the thickness. We thus infer that the magnetization switching for the thicker underlayer films is due to the spin Hall torque at the HM|CoFeB interface, whereas the switching for the thin underlayer films is dominated by spin transfer torque within the CoFeB layer[24, 25].

## C. In-plane field dependence of the threshold current

To gain insight into the respective roles of the spin transfer torques and the spin Hall torques for driving magnetization reversal, we have studied the threshold current as a function of in-plane external fields. Figure 3 shows $J_N^C$ as a function of in-plane field along $x$ ($H_X$) and $y$ ($H_Y$) for films with thin and thick TaN underlayer films. The squares and circles represent initial magnetization pointing along $+z$ and $–z$, respectively.

For the thin underlayer films (Figs. 3(a) and 3(b)), $J_N^C$ is symmetric with respect to the in-plane field. Figure 3(a) shows that magnetization switching is assisted by $+H_X$ for positive current when the initial magnetization direction points along $–z$. $J_N^C$ tends to saturate as the magnitude of $H_X$ is increased. In contrast, Fig. 3(b) shows that the threshold current is strongly influenced by $H_Y$ within the same field range: the difference in $J_N^C$ for initial magnetization pointing along $+z$ and $–z$ increases with increasing $|H_Y|$. For these films, the current-induced effective field due to the spin Hall effect is small and we can therefore assume that the STT (current through the magnetic layer) plays the dominant role in the magnetization reversal process. Theoretically, it has been reported that STT can cause spin wave instabilities that consequently result in domain wall nucleation, or partial magnetization reversal, when large enough current is applied[25, 26]. In such cases, the threshold current needed to cause magnetization switching does not, as a first approximation, depend on small (compared to the anisotropy field) in-plane applied field[27]. Further study is required to identify the origin of the in-plane field dependence.

For the thicker underlayer films, the threshold current density exhibits a different in-plane field dependence. As described above, $J_N^C$ is different for initial magnetization pointing along $+z$ and $-z$ in the absence of external field. This difference in $J_N^C$, for a given current direction, reverses when a small in-plane field directed along the $+x$ direction is applied (Fig. 3(c)). The field needed to match $J_N^C$ for positive and negative currents, termed the offset field ($H_X^*$) hereafter, is ~20-25 Oe for the sample shown in Fig. 3(c). The offset field $H_X^*$ is plotted as a function of the TaN underlayer thickness in Fig. 4(d). We find that $H_X^*$ increases with the TaN underlayer thickness: the reason behind this will be discussed in section III in connection with the ratio of the field-like ($\Delta H_Y$, Fig. 4(a)) and damping-like ($\Delta H_X$, Fig. 4(b)) components of the spin Hall effective field, shown in Fig. 4(c).

Previously, it has been reported that a non-zero $H_X$ is needed to switch the magnetization directed along the film normal with in-plane current[3, 8, 18]. Here, owing to the non-zero $H_X^*$, magnetization switching can be triggered at zero magnetic field. Note that the threshold current dependence on $H_X$ is consistent with the negative spin hall angle of the underlayer[3, 21]: the threshold current is smaller when the direction of $H_X + H_X^*$ matches that of the damping like component of the spin Hall effective field compared to the opposite case. The damping like component of the spin Hall effective points along the $-x$ direction for positive current and magnetization pointing along +z: it points in the opposite direction if the current or the magnetization direction is reversed; see Fig. 4(b).

For in-plane field ($H_Y$) applied perpendicularly to the current flow, $J_N^C$ is found to vary more or less linearly with $H_Y$ (Fig. 3(d)). The dependence of $J_N^C$ on $H_Y$ is compared to model calculations in section III to discuss its relationship with the sign of the field-like spin Hall torque.

### D. Dependence on the film deposition conditions

The zero field switching found here indicates that the symmetry of the system is broken for the thick underlayer films. We find that the symmetry breaking factor arises during the film deposition (sputtering) process. Figure 5(a) shows schematic of the inside of sputtering chamber with focus on the relation between the substrate position and the sputtering target. The same coordinate axes shown in Fig. 1(a) are illustrated in Fig. 5(a) for reference. Three substrates are placed for film deposition and we find that the asymmetry in the switching with current changes depending on the position of the substrate. Figure 5(b) and 5(c) show Kerr images after voltage pulses are applied to the wire when the initial magnetization direction is set along −z (the films has 3.6 nm thick TaN underlayer). When the substrate is positioned along the +y direction, denoted as "Left" in Fig. 5(a), the switching probability (i.e. the area with brighter contrast) is larger for negative current (Fig. 5(b)). This asymmetry is the same with that shown in Figs. 1-3. In contrast, when the substrate is placed along the −y direction (referred to as the "Right" position in Fig. 5(a)), the asymmetry reverses: the switching probability is now larger for the positive current. The pulse amplitude dependence of the switching probability is shown in Fig.

5(d), which clearly shows the difference in the asymmetry. We have also studied current induced magnetization switching for wires whose long axis is directed along the *y*-axis (Fig. 6). In such case, we find little difference in the switching current for positive/negative currents and the initial magnetization along ±*z*.

The asymmetric magnetization switching is also found in other heavy metal underlayer films (Hf and W). As shown in Fig. 7, the asymmetry of the switching with respect to the current flow direction and the initial magnetization direction is the same for all underlayer films as long as the position of the substrate is kept same. Note that the sign of the spin Hall angle for the heavy metals used here is the same whereas the Dzyaloshinskii-Moriya interaction (DMI)[28, 29] at the underlayer|CoFeB layer interface changes its sign between Hf and W[21].

## E. The effect of the out of plane field

A non-zero out of plane magnetic field can introduce difference in the switching probability for initial magnetization pointing along +*z* and −*z*. Figure 8 shows the pulse amplitude dependence of the switching probability when the out of plane field ($H_Z$) is varied. As evident, the switching probability is larger for both current flow directions when $H_Z$ assists the switching process, i.e. when $H_Z$ is pointing opposite to the initial magnetization direction. However, these results show that $H_Z$ by itself cannot induce difference in the switching for positive and negative currents. The maximum residual field from the electromagnet at the sample position is ~1 Oe.

## F. Pulse length dependence and repeated switching measurements

The magnetization switching observed here may be influenced by subsequent motion of nucleated domain walls driven by current[30, 31]. To study whether the asymmetry of $J_N^C$ with the current and initial magnetization directions is due to motion of domain walls, we have studied the pulse length dependence of $J_N^C$. If any subsequent domain wall motion is causing the asymmetry, such effect should diminish when the pulse length is reduced since the distance the domain wall travels will also decrease. Figure 9(a) shows $J_N^C$ as a function of pulse length ($t_P$) for the device shown in Fig. 1(d), in which we consider spin Hall torque is responsible for the switching. A pulse train consisting of five $t_P$ ns-long-pulses, each separated by 10 ms, is applied. The difference in $J_N^C$ for positive and negative currents as well as that for initial magnetization pointing along $+z$ and $-z$ remains the same even for pulse length of 10 ns. We have observed such asymmetry in other devices for pulse length as small as 5 ns. Thus these results show that the asymmetry is predominantly caused by the nucleation process and not the subsequent domain wall motion.

In Fig. 9(b), we show that the switching process can be deterministic even in the absence of magnetic field. A pulse train consisting of five 100 ns-long-pulse is used for each "pulse" shown in the top panel. The sign of the pulse train is altered each time. We have chosen the same device shown in Fig. 1(d) in which the asymmetry is large so that full switching of magnetization takes place upon the pulse application (if the asymmetry is small, it is difficult to reverse the entire area of the wire with a single pulse). The middle and bottom panels of Fig. 9(b) show the variation of the magnetic state, inferred from the Kerr images, with successive pulse application. The state at the beginning (i.e. "Iteration 0") has different orientation for the middle and bottom panels. When the magnetization is pointing along $+z$ ($-z$), positive (negative) current can trigger magnetization reversal. Full switching of the wire magnetization is observed when appropriate pulse is applied. When a "wrong" pulse is applied, as shown at "Iteration 1" in the bottom panel, we do not find random nucleation due to thermal activation, thus showing the robustness of this switching scheme.

## III. Model calculations

## A. Macrospin model

To gain insight of the asymmetric magnetization switching with current and the in-plane field dependence of $J_N^C$, we show results from a model calculation using the Landau-Lifshitz-Gilbert (LLG) equation. We find that if we assume an uniaxial magnetic anisotropy that is tilted away from the normal of the film plane, a mechanism first suggested in Ref. [20], many of our experimental results can be explained. Similar results can be obtained if a unidirectional anisotropy pointing along the wire's long axis is assumed. However, with this assumption, $H_X^*$ will simply be defined by the unidirectional anisotropy field and it is difficult to explain some of the experimental results, for example, the TaN underlayer thickness dependence of $H_X^*$ (Fig. 4(d)). The LLG equation that includes the spin Hall torques reads:

$$\frac{\partial \hat{m}}{\partial t} = -\gamma \hat{m} \times \left( \vec{H}_K + \vec{H}_{EXT} + a_J (\hat{m} \times \hat{p}) + b_J \hat{p} \right) + \alpha \hat{m} \times \frac{\partial \hat{m}}{\partial t} \qquad (1)$$

where $\hat{m}$ is a unit vector representing the magnetization direction, $t$ is time, $\gamma$ is the gyromagnetic ratio and $\alpha$ is the Gilbert damping parameter. $\vec{H}_K$ and $\vec{H}_{EXT}$ represent the uniaxial anisotropy field and the external magnetic field, respectively. We set the axis of the uniaxial anisotropy field to be oriented along a unit vector $\hat{k}$, i.e. $\vec{H}_K = H_K (\hat{m} \cdot \hat{k}) \hat{k}$. The coordinate system employed in the calculations is the same as that shown in Fig. 1(a).

The effect of current is coded in the parameters $a_J$ and $b_J$. $a_J$ is the damping-like component of the spin Hall effective field[1,2] whereas $b_J$ corresponds to the field-like component[32]. We assume that $a_J$ and $b_J$ are proportional to current that flows through the wire. Unit vector $\hat{p}$ represents the spin direction of the electrons that impinge upon the magnetic layer (FM) generated within the heavy metal layer (HM) via the spin Hall effect. Positive current corresponds to current flow along the +x direction. For positive current, we set $\hat{p} = (0,1,0)$ as this represents the spin direction of the electrons entering the CoFeB layer via the spin Hall effect in heavy metal layers with negative spin Hall angle such as Ta and W. We vary $a_J$ and $b_J$ to study the effect of current. Current and field are applied to the system and the resulting equilibrium magnetization direction is calculated 100 ns after the current/field application. In order to cause magnetization switching within reasonable values of $a_J$, we use a reduced uniaxial anisotropy field[18], i.e. $H_K \sim 530$ Oe.

Figure 10 shows results of model calculations when the uniaxial anisotropy axis is tilted in the yz plane, i.e. $\hat{k} = (0 \quad \sin\beta \quad \cos\beta)$. Here we set the tilt angle $\beta$ to be 2 degree away from the z axis. Figures 10(a) and 10(b) show the z-component of magnetization as a function of $a_J$. The sign of $b_J$ is opposite for Figs. 10(a) and 10(b). As evident, the z-component of the magnetization ($m_Z$) rotates toward the film plane as $a_J$ is increased. In many cases, we find an abrupt transition of the magnetic state from the film normal to the film plane. Once the magnetization points along the film plane, it can move back to its original direction or it can move to the opposite side of the z-axis, resulting in magnetization switching, after the current is turned off due to thermal activation. We thus define the threshold $a_J$ ($a_J^C$) as the minimum $a_J$ needed to cause the absolute value of $m_Z$ to be less than 0.15: this value is justified by micromagnetic simulations shown in the next section.

Note that in some cases (e.g. Fig. 10(b)), we find the equilibrium $m_Z$ during the current application jumps to the equilibrium position of the other branch (i.e. opposite to the initial direction). This indicates deterministic switching of the magnetization, not the probabilistic switching as described above: the direction of switching is predefined during the current application. Interestingly, such deterministic switching will diminish as $a_J$ is further increased since the equilibrium $m_Z$ during the current application favors the direction along the film plane, resulting in the probabilistic switching. Such drop in the switching probability with increasing current density is also found in experiments (see e.g. Fig. 5d).

Figures 10(c-e) show the in-plane field ($H_X$) dependence of $a_J^C$ when the field-like component ($b_J$) is varied. The asymmetric magnetization switching with non-zero $H_X^*$ is reproduced here with the tilt angle $\beta$ set to 2 deg. The sign of $H_X^*$ is independent of the size and sign of $b_J$. The negative $H_X^*$ shown in Fig. 10(c-e) is found experimentally in samples deposited in the "Right" position defined in Fig. 5(a). Due to the non-zero tilting of the anisotropy axis that breaks the symmetry of the system, $a_J^C$ is different for positive and negative currents for a given initial magnetization direction.

Interestingly, $H_X^*$ not only depends on the tilting angle ($\beta$), but also on the relative size of the field-like and damping-like components of the spin Hall torque, that is, the size of $b_J/a_J$. The model shows that $H_X^*$ exhibits a complex dependence on $b_J/a_J$: $H_X^*$ takes a maximum when $b_J = -a_J$ and it drops as $|b_J|$ further increases. Experimentally, we have previously studied the underlayer thickness dependence of the spin Hall torque using the harmonic Hall measurements[21, 23]: Figs. 4(a) and 4(b) show the field-like ($\Delta H_Y = b_J \hat{p} \cdot \hat{y}$) and the damping-like ($\Delta H_X = a_J (\hat{m} \times \hat{p}) \cdot \hat{x}$) components of the spin Hall effective field, respectively. The ratio of the two components, $-\Delta H_Y/\Delta H_X$ is equal to $b_J/a_J$ and is plotted in Fig. 4(c). Although the number of data is limited, the thickness dependence of $H_X^*$, plotted in Fig. 4(d), shows that it more or less scales with $b_J/a_J$. These results show that $H_X^*$ is not only a function of the sample position during the sputtering but also dependent on the characteristics of the spin Hall torque. The detailed difference between the model calculations and the experimental results require further thorough study of $H_X^*$.

We have also studied the in-plane field dependence of $a_J^C$ when the direction of the uniaxial anisotropy axis ($\hat{k}$) is varied. When the tilt direction is inverted in the $yz$ plane, i.e. $\hat{k} = (0 \ -\sin\beta \ \cos\beta)$, the sign of $H_X^*$ reverses. Experimentally, $H_X^*$ changes its sign when the position of the substrate during the sputtering is changed, as shown in Fig. 5. These results indicate that the tilt angle depends on the substrate position. $H_X^*$ is zero and the asymmetric magnetization switching disappears when the tilt direction is set along the $xz$ plane, i.e. $\hat{k} = (\sin\beta \ 0 \ \cos\beta)$. This is in agreement with the results shown in Fig. 6, where the asymmetry diminishes when the wire's long axis is oriented along the tilt direction (i.e. along the $y$-axis).

It is somewhat counterintuitive to understand why an offset field in the $x$-direction ($H_X^*$) emerges (e.g. Fig. 3c) when the uniaxial anisotropy field is tilted along the $yz$ plane with a tilt angle $\beta$. One way to understand this is to view the incoming spins diffusing from the HM layer into the magnetic layer in the frame along the tilted anisotropy axis. The polarization $\hat{p}$ directed along the $+y$ direction in the lab frame has to be changed to $\hat{p}' = (0 \ \cos\beta \ \sin\beta)$ in a rotated frame defined by the tilted anisotropy axis. The polarization possesses a non-zero component (i.e. $\sin\beta$) along the easy axis that can cause the difference in the switching current for opposite initial magnetization directions and current flow directions, similar to conventional spin transfer torque switching of parallel/antiparallel magnetization. The tilted anisotropy axis thus breaks the symmetry along the *z*-direction, which in turn manifests itself as an offset field in the *x*-direction.

The bottom panels of Fig. 10 show the $H_Y$ dependence of $a_J^C$ for different values of $b_J$. When the sign of $b_J$ is opposite to that of $a_J$ (Fig. 10f), $a_J^C$ monotonically varies with $H_Y$. This is in agreement with the experimental results shown in Fig. 3(d). The slope of $a_J^C$ versus $H_Y$ around zero field changes as the size and sign of $b_J$ is varied (Fig. 10(g,h)). These results show that the slope of $J_N^C$ vs. $H_Y$ around zero field roughly gives the sign of the field-like torque ($b_J$).

## B. Micromagnetic simulations

We have performed micromagnetic simulations to validate the macrospin model used to describe the experimental results. The micromagnetic code "mumax"[33] is used for the simulations. The magnetic parameters used in the simulations are described in the caption of Fig. 11: the parameters are chosen so that the magnetic anisotropy is the same with that used in the macrospin calculations (Fig. 10). The definition of the coordinate axis is drawn in the inset to Fig. 11(a). The anisotropy axis is tilted along the *yz* plane by 1 deg. Here we use $b_J=a_J$ since this condition gives the largest difference in the switching current for opposite initial magnetization directions at zero field in the macrospin model.

The procedure of simulation is the following: a temperature pulse of 700 K and duration of 0.2 ns is first applied to an uniform magnetic state to mimic the experimental condition, i.e. thermal agitation of the magnetization. A pulse current of 1 ns is applied to study the magnetic state during the current application. The current flows along the +*x* direction. We have checked the effect of the current pulse length and find that 1 ns is long enough to study the switching process in most cases. The current is then turned off and the system is relaxed to study the final state of the magnetization. The equilibrium magnetic state during the current application for positive current is plotted in Fig. 11(a) for initial magnetization states along ~+z (black squares) and ~−z (red circles). The results are equivalent to those of macrospin calculations, Fig. 10(b). As evident, the equilibrium state favors to point along ~+z for this current direction. When the initial magnetization points along ~−z, there is a critical $a_J$ above which magnetization switches its direction during the current application. This is equivalent to the deterministic switching found in the macrospin calculations. When the current is further increased the magnetization falls closer to the film plane.

The switching probability after the current is turned off and the system is relaxed is shown in Fig. 11(b) as a function of $a_J$ for both initial magnetic states. The switching probability is obtained from the area of the element that switched divided by the whole area, similar to the method used in the experiments. For initial magnetization pointing along –z, we find full (i.e. deterministic) switching of the magnetization above $a_J$~400 Oe. For the opposite initial magnetic state (along +z), the switching probability saturates at ~0.5 for large $a_J$. Note that probability~0.5 corresponds to a multi-domain state as shown in the inset to Fig. 11(a). We find that if $|m_Z|$ during the current application is less than ~0.13, denoted by the blue dashed line in Fig. 11(a), domain walls can nucleate during the relaxation process and the final state is a multi-domain state. In other words, if $|m_Z|$ is larger ~0.13, the final state possesses the same magnetization configuration with the initial magnetic state unless the deterministic switching occurs. This justifies our assumption on using $m_Z$=0.15 for calculating the threshold $a_J$ for magnetization switching in the macrospin model. The features found in the simulations are in agreement with experiments, where full switching of magnetization is observed only in one of the starting condition for a given current direction, while the other only produces a multi-domain state, i.e. partial magnetization switching.

## IV. Discussion

Besides from the tilted uniaxial anisotropy which we consider breaks the symmetry in our system, other factors can also cause the asymmetry in magnetization reversal with current. Recently, it has been reported that a gradient in the magnetic anisotropy across the wafer can break the symmetry and enable zero field switching. Here, as the underlayer thickness is varied along the $x$ direction, it creates a gradient in the magnetic anisotropy and the saturation magnetization across the wafer. This is in contrast to Ref. 19 in which the gradient is created along the $y$ axis in our definition (see Fig. 5(a)). We thus consider that the effect of the out of plane field-like torque proposed in Ref. 19 may be minor here.

The asymmetric shape of the patterned wire (Fig. 1(a)), where the right side of the wire is connected to a region with low magnetic anisotropy due to prior etching of half the MgO layer and the Ta capping layer before the Ta|Au pad formation, can result in preferential current induced injection of domain walls from the right side of the wire[34]. We have thus tested symmetric structures with large pads attached to both sides of the wire and have found that the asymmetry is not altered.

The DMI can play a role in the nucleation process[31, 35, 36]. As reported in Ref. 36, for an uniform initial magnetization state, the DMI is relevant near the edge of the wire where the magnetization is tilted. We find little evidence of nucleation events taking place preferentially from the edges of the wire for many of the films studied here. One exception is the W underlayer films, where we find preferential nucleation from the edges when a relatively large (a few hundred Oersteds) in-plane field along the wire's long axis ($H_X$) is applied. However, the nucleated region is limited to the edge of the wire (near the Ta|Au electrodes) and cannot explain the full reversal that occurs within the wire. As shown in Fig. 7, the asymmetric magnetization switching with current occurs in a similar fashion for the Hf and W underlayer films, which possess opposite sign of the interface DMI[21]. We thus infer that the DMI is not the main source of the asymmetric switching.

## V. Conclusion

In summary, we have studied current driven magnetization switching in perpendicularly magnetized CoFeB|MgO heterostructures with heavy metal underlayers (TaN). The threshold current needed to reverse the magnetization direction is studied as a function of film structure, pulse amplitude, pulse length and in-plane magnetic field. From the in-plane magnetic field dependence we find that magnetization switching takes place via spin transfer torque within the CoFeB layer when the underlayer thickness is small, whereas the switching occurs due to spin Hall torque for thicker underlayer films. For spin Hall torque driven magnetization reversal, the threshold current is different for positive and negative currents as well as the initial magnetization directions (pointing along +z or –z). We attribute such asymmetry of the switching current to a tilting of the uniaxial anisotropy axis, away from the normal of the film plane, which develops during the film deposition process (sputtering). The asymmetry depends on the relative position of the substrate and the center of the sputtering target, suggesting an extrinsic origin.

Just a few degrees of the tilting can break the symmetry to enable zero field switching of perpendicularly magnetized thin films using in-plane current.

**Acknowledgements**

We thank G. Tatara for helpful comments on the experimental results and J. Kim and T. Devolder for technical support. This work was partly supported by MEXT R & D Next-Generation Information Technology and the Grant-in-Aid for Young Scientists (A).

# References

1　J. C. Slonczewski, J. Magn. Magn. Mater. **159**, L1 (1996).
2　L. Berger, Phys. Rev. B **54**, 9353 (1996).
3　L. Liu, C.-F. Pai, Y. Li, H. W. Tseng, D. C. Ralph, and R. A. Buhrman, Science **336**, 555 (2012).
4　C. F. Pai, L. Q. Liu, Y. Li, H. W. Tseng, D. C. Ralph, and R. A. Buhrman, Appl. Phys. Lett. **101** (2012).
5　M. Yamanouchi, L. Chen, J. Kim, M. Hayashi, H. Sato, S. Fukami, S. Ikeda, F. Matsukura, and H. Ohno, Appl. Phys. Lett. **102**, 212408 (2013).
6　K. Garello, C. O. Avci, I. M. Miron, M. Baumgartner, A. Ghosh, S. Auffret, O. Boulle, G. Gaudin, and P. Gambardella, arXiv:1310.5586 (2013).
7　M. Cubukcu, O. Boulle, M. Drouard, K. Garello, C. Onur Avci, I. Mihai Miron, J. Langer, B. Ocker, P. Gambardella, and G. Gaudin, Appl. Phys. Lett. **104** (2014).
8　I. M. Miron, K. Garello, G. Gaudin, P. J. Zermatten, M. V. Costache, S. Auffret, S. Bandiera, B. Rodmacq, A. Schuhl, and P. Gambardella, Nature **476**, 189 (2011).
9　M. I. Dyakonov and V. I. Perel, Phys. Lett. A **35**, 459 (1971).
10　X. Wang and A. Manchon, Phys. Rev. Lett. **108**, 117201 (2012).
11　K. W. Kim, S. M. Seo, J. Ryu, K. J. Lee, and H. W. Lee, Phys. Rev. B **85** (2012).
12　P. M. Haney, H. W. Lee, K. J. Lee, A. Manchon, and M. D. Stiles, Phys. Rev. B **87** (2013).
13　J. Kim, J. Sinha, S. Mitani, M. Hayashi, S. Takahashi, S. Maekawa, M. Yamanouchi, and H. Ohno, Phys. Rev. B **89**, 174424 (2014).
14　C.-F. Pai, Y. Ou, D. C. Ralph, and R. A. Buhrman, arXiv:1411.3379 (2014).
15　S. Ikeda, K. Miura, H. Yamamoto, K. Mizunuma, H. D. Gan, M. Endo, S. Kanai, J. Hayakawa, F. Matsukura, and H. Ohno, Nat. Mater. **9**, 721 (2010).
16　D. C. Worledge, G. Hu, D. W. Abraham, J. Z. Sun, P. L. Trouilloud, J. Nowak, S. Brown, M. C. Gaidis, E. J. O'Sullivan, and R. P. Robertazzi, Appl. Phys. Lett. **98**, 022501 (2011).
17　K. S. Lee, S. W. Lee, B. C. Min, and K. J. Lee, Appl. Phys. Lett. **104** (2014).
18　L. Q. Liu, O. J. Lee, T. J. Gudmundsen, D. C. Ralph, and R. A. Buhrman, Phys. Rev. Lett. **109** (2012).
19　G. Q. Yu, P. Upadhyaya, Y. B. Fan, J. G. Alzate, W. J. Jiang, K. L. Wong, S. Takei, S. A. Bender, L. T. Chang, Y. Jiang, M. R. Lang, J. S. Tang, Y. Wang, Y. Tserkovnyak, P. K. Amiri, and K. L. Wang, Nat. Nanotechnol. **9**, 548 (2014).
20　L. You, O. Lee, D. Bhowmik, D. Labanowski, J. Hong, J. Bokor, and S. Salahuddin, arXiv: 1409.0620 (2014).
21　J. Torrejon, J. Kim, J. Sinha, S. Mitani, M. Hayashi, M. Yamanouchi, and H. Ohno, Nature Comm. **5**, 4655 (2014).
22　J. Sinha, M. Hayashi, A. J. Kellock, S. Fukami, M. Yamanouchi, M. Sato, S. Ikeda, S. Mitani, S. H. Yang, S. S. P. Parkin, and H. Ohno, Appl. Phys. Lett. **102** (2013).
23　J. Kim, J. Sinha, M. Hayashi, M. Yamanouchi, S. Fukami, T. Suzuki, S. Mitani, and H. Ohno, Nat. Mater. **12**, 240 (2013).
24　K. J. Lee, A. Deac, O. Redon, J. P. Nozieres, and B. Dieny, Nat. Mater. **3**, 877 (2004).
25　J. Shibata, G. Tatara, and H. Kohno, Phys. Rev. Lett. **94**, 076601 (2005).
26　Z. Li and S. Zhang, Phys. Rev. Lett. **92**, 207203 (2004).
27　G. Tatara, private communications.
28　I. E. Dzyaloshinskii, Sov. Phys. JETP **5**, 1259 (1957).
29　T. Moriya, Phys. Rev. **120**, 91 (1960).

[30] O. J. Lee, L. Q. Liu, C. F. Pai, Y. Li, H. W. Tseng, P. G. Gowtham, J. P. Park, D. C. Ralph, and R. A. Buhrman, Phys. Rev. B **89** (2014).
[31] G. Q. Yu, P. Upadhyaya, K. L. Wong, W. J. Jiang, J. G. Alzate, J. S. Tang, P. K. Amiri, and K. L. Wang, Phys. Rev. B **89** (2014).
[32] S. Zhang, P. M. Levy, and A. Fert, Phys. Rev. Lett. **88**, 236601 (2002).
[33] A. Vansteenkiste, J. Leliaert, M. Dvornik, M. Helsen, F. Garcia-Sanchez, and B. Van Waeyenberge, Aip Advances **4** (2014).
[34] M. Hayashi, M. Yamanouchi, S. Fukami, J. Sinha, S. Mitani, and H. Ohno, Appl. Phys. Lett. **100**, 192411 (2012).
[35] N. Perez, E. Martinez, L. Torres, S. H. Woo, S. Emori, and G. S. D. Beach, Appl. Phys. Lett. **104** (2014).
[36] S. Pizzini, J. Vogel, S. Rohart, L. D. Buda-Prejbeanu, E. Jue, O. Boulle, I. M. Miron, C. K. Safeer, S. Auffret, G. Gaudin, and A. Thiaville, Phys. Rev. Lett. **113** (2014).

**Figure captions:**

**Figure 1.** (a) Exemplary optical microscopy image of the wire used to study current-induced magnetization switching. The dark regions indicate the magnetic film whereas the yellow regions represent the Ta|Au electrodes. A pulse generator is connected to the left electrode. (b) Out of plane hysteresis loops measured using Kerr microscopy for Sub.|$d$ TaN|1 CoFeB|2 MgO|1 Ta: $d$=3.6 nm (red circles) and $d$=0.5 nm (black squares). (c-d) Magnetization switching probability as a function of pulse amplitude for initial magnetization configurations pointing along +$z$ (black squares) and –$z$ (red circles) for the two devices shown in (b). Positive and negative probability corresponds to initial magnetization direction pointing along +$z$ and –$z$, respectively. A pulse train consisting of five 100 ns-long-pulses is applied. Representative Kerr images captured after the application of the maximum amplitude pulse (both positive and negative voltages) are included at the corresponding corners of each panel. Results are from substrates placed in the "Left" position defined in Fig. 5(a).

**Figure 2.** Threshold current density ($J_N^C$) as a function of TaN underlayer thickness. The initial magnetization direction points along –$z$ (a) and +$z$ (b). Solid and open symbols represent positive and negative $J_N^C$, respectively. A pulse train consisting of five 100 ns-long-pulses is applied. Results are from substrates placed in the "Left" position defined in Fig. 5(a).

**Figure 3.** In-plane field dependence of the threshold current density ($J_N^C$). The field direction is along (a,c) and transverse to (b,d) the current flow. The underlayer is TaN: its thickness is 0.5 nm (a,b) and 6.6 nm (c,d). Black squares and red circles represent initial magnetization direction

along +z and −z, respectively. A pulse train consisting of five 100 ns-long-pulses is applied. Results are from substrates placed in the "Left" position defined in Fig. 5(a).

**Fig. 4**. The field-like ($\Delta H_Y$) (a) and the damping-like ($\Delta H_X$) (b) components of the current induced effective field plotted against the TaN underlayer thickness (source: Ref. [21]). Black squares and red circles correspond to magnetization directed along +z and −z, respectively. The effective field is normalized by the current density $J_N$ that flows through the TaN layer. (c) Ratio of the field-like component to the damping like component, $-\Delta H_Y/\Delta H_X$, plotted against the TaN underlayer thickness. A minus sign is multiplied so that $-\Delta H_Y/\Delta H_X$ is equal to $b_J/a_J$ defined in Eq. (1). (d) TaN thickness dependence of the offset field $H_X^*$. Solid and open symbols correspond to $H_X^*$ estimated using positive and negative currents.

**Figure 5.** (a) Schematic illustration of inside the sputtering chamber where the relative position of the substrates and the target is shown. Three ~1×1 cm² square substrates, separated by ~0.15 cm along the y direction, are placed ~10 cm away from the target. (b,c) Kerr images after application of ±32 V voltage pulses for devices made of the same film (Sub.|3.6 nm TaN|1 CoFeB|2 MgO|1 Ta) but the substrates are placed at different positions: (b) "Left" position and (c) "Right" position defined in (a). The top and bottom images correspond to images when positive and negative voltage pulses are applied, respectively. (d) Magnetization switching probability as a function of pulse amplitude for the two devices shown in (b) and (c). The initial magnetization direction points along −z. A pulse train consisting of five 100 ns-long-pulses is applied for (b-d).

**Fig. 6**. Pulse amplitude dependence of magnetization switching probability for Sub.|2.9 TaN|1 CoFeB|2 MgO|1 Ta (units in nm). The patterned wires' long axis is directed along *x* (a) and *y* (b). Results are from substrates placed in the "Left" position defined in Fig. 5(a). Positive and negative probability corresponds to initial magnetization direction pointing along +z and –z, respectively.

**Fig. 7.** Magnetization switching probability as a function of pulse amplitude for initial magnetization configurations pointing along +z (black squares) and –z (red circles) for devices with different heavy metal underlayers. The films are Sub.|*d* X|1 CoFeB|2 MgO|1 (units in nanometers), with X=5.9 nm Hf (a) and 3.1 nm W (b). A pulse train consisting of five 100 ns-long-pulses is applied. Positive and negative probability corresponds to initial magnetization direction pointing along +z and –z, respectively. Results are from substrates placed in the "Left" position defined in Fig. 5(a).

**Fig. 8**. Magnetization switching probability as a function of pulse amplitude for Sub.|2.9 TaN|1 CoFeB|2 MgO|1 Ta (units in nm). The out of plane field $H_Z$ is varied: $H_Z$~−5 (a), ~0 (b) and (c) ~+5 Oe. Positive and negative probability corresponds to initial magnetization direction pointing along +z and –z, respectively. Results are from substrates placed in the "Left" position defined in Fig. 5(a).

**Figure 9.** (a) Threshold current density ($J_N^C$) vs. pulse length (*t*) at zero external field for Sub.|3.6 nm TaN|1 CoFeB|2 MgO|1 Ta. A pulse train consisting of five *t* ns-long-pulses, each separated by 10 ms, is applied. Black squares and red circles show $J_N^C$ when the initial magnetization direction is along +z and –z, respectively. (b) Sequences of voltage pulses applied to the wire (top panel) and the resulting Kerr contrast ($\Delta I$) calculated from the Kerr images. The corresponding magnetic state (1: along +z, -1: along –z) is shown in the right axis. A pulse train consisting of five 100 ns-long pulses, each separated by 10 ms, is applied at each pulse shown in the top panel. Middle and bottom panels of (b) show changes in the Kerr contrast for initial magnetization pointing along +z and –z at the beginning of the sequence, respectively.

**Fig. 10**. (a,b) *z*-component of the equilibrium magnetization when current and in-plane magnetic field are turned on, plotted as a function of $a_J$, the damping like component of the spin Hall torque. The field-like component of the spin Hall torque $b_J$ is set to $-a_J$ (a) and $a_J$ (b). The horizontal blue dashed lines indicate $|m_Z|=0.15$, which is used to define $a_J^C$. (c-h) $a_J^C$ as a function of $H_X$ (c-e) and $H_Y$ (f-h). The field-like component $b_J$ is varied: $b_J=-a_J$ (c,f), $b_J=0$ (d,g) and $b_J=a_J$ (e,h). For all plots, black squares and red circles represent calculation results when the initial magnetization direction points along +z and –z, respectively. $H_K$=528 Oe, $\alpha$=0.05, the uniaxial anisotropy axis (direction defined by a unit vector $\hat{k}$) is tilted 2 deg toward the *y*-axis, i.e. $\hat{k}$=(0, sin$\beta$, cos$\beta$) with $\beta$=2 deg.

**Fig 11**. Micromagnetic simulations of spin Hall torque driven magnetization switching. (a,b) $a_J$ (the damping like component of the spin Hall torque) dependence of the *z*-component of magnetization at the end of 1 ns pulse (a) and the switching probability calculated from the magnetic state 20 ns after the pulse is turned off (b). Black squares and red circles represent initial magnetization along $+z$ and $-z$, respectively. Parameters used are: saturation magnetization $M_S$=1250 emu/cm$^3$, exchange constant $A$=3.1×10$^{-6}$ erg/cm, uniaxial magnetic anisotropy energy $K$=10.15×10$^{-6}$ erg/cm$^3$, Gilbert damping $\alpha$=0.05 and the field-like component of the spin Hall torque $b_J$=$a_J$. The dimension of the simulated element is 2000×500×1 nm$^3$ with a discretization cell of ~2×2×1 nm$^3$. The anisotropy axis is tilted along the $yz$ plane by 1 deg. Inset to (b): simulated magnetization image 20 ns after a pulse of $a_J$=368.6 Oe is turned off: the initial magnetization is along $-z$.

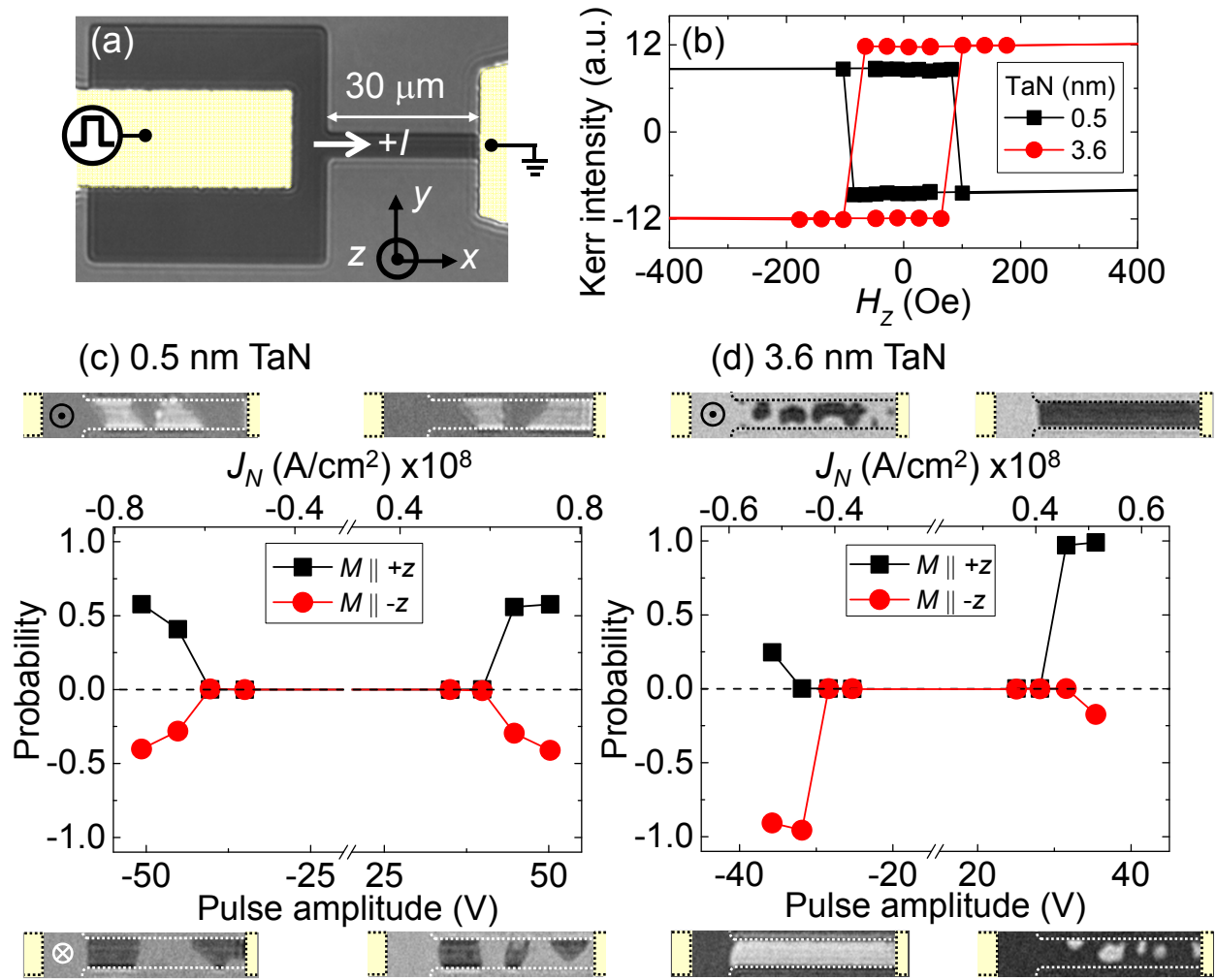

Fig. 1

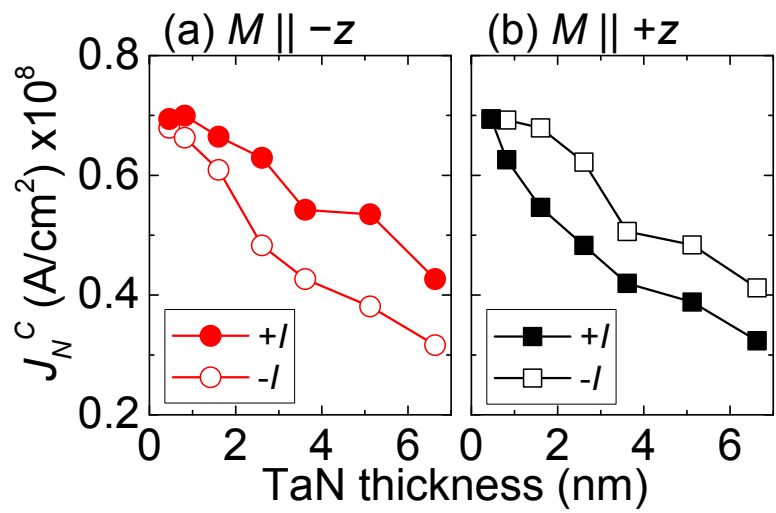

Fig. 2

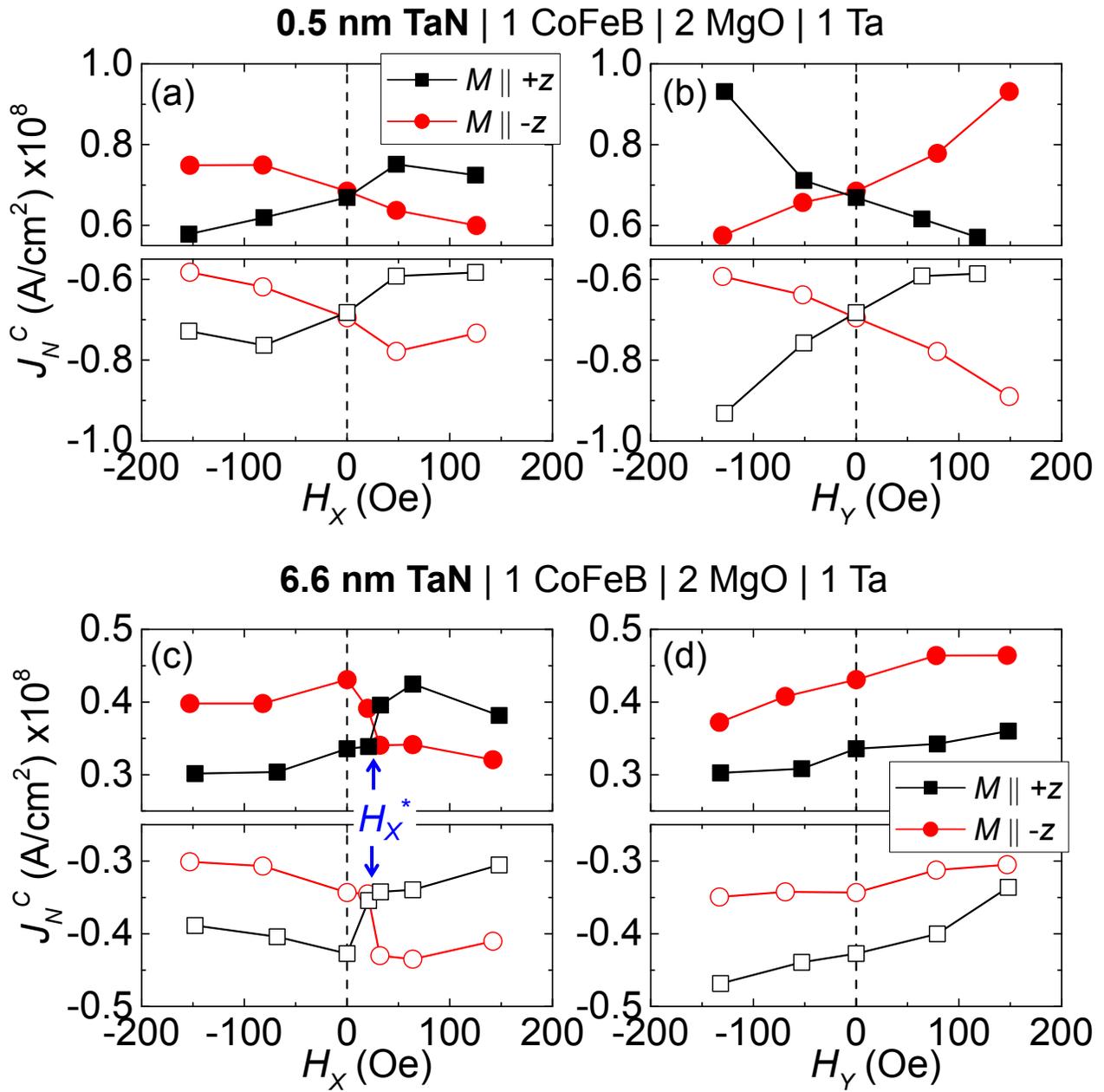

Fig. 3

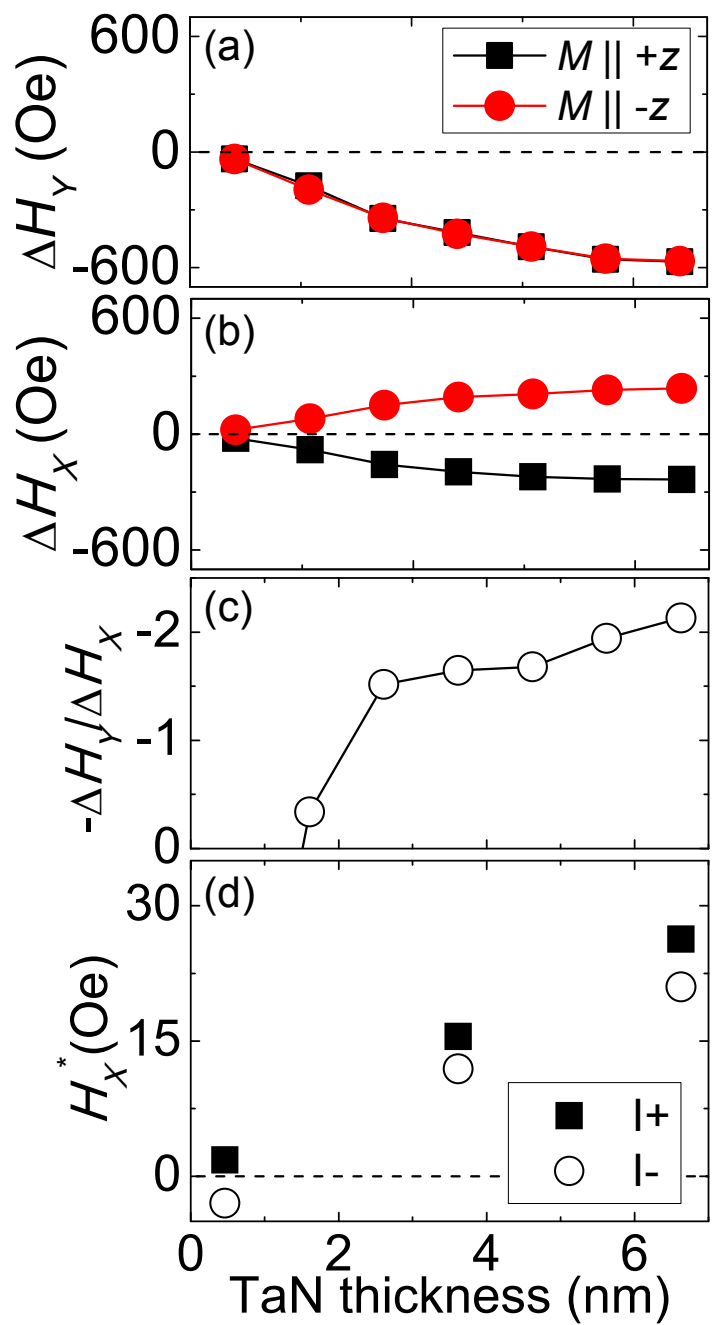

Fig. 4

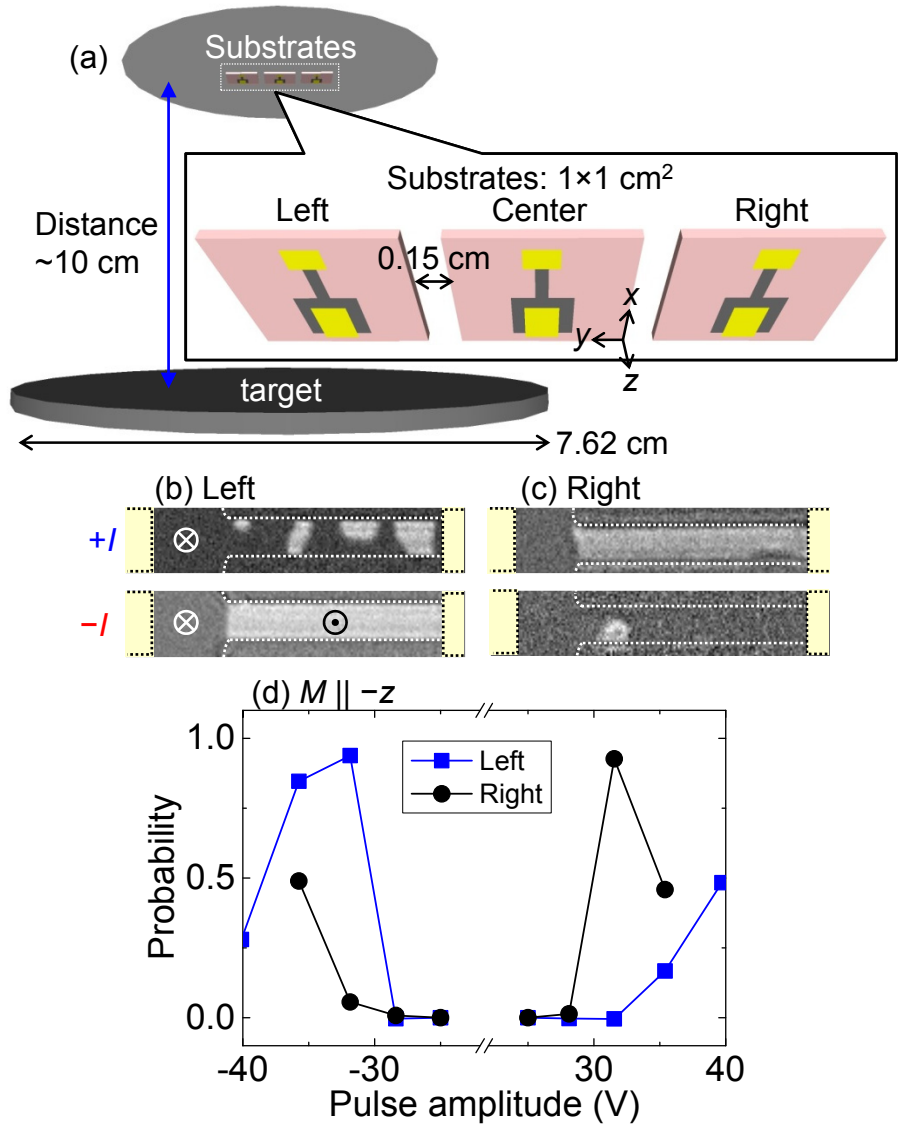

Fig. 5

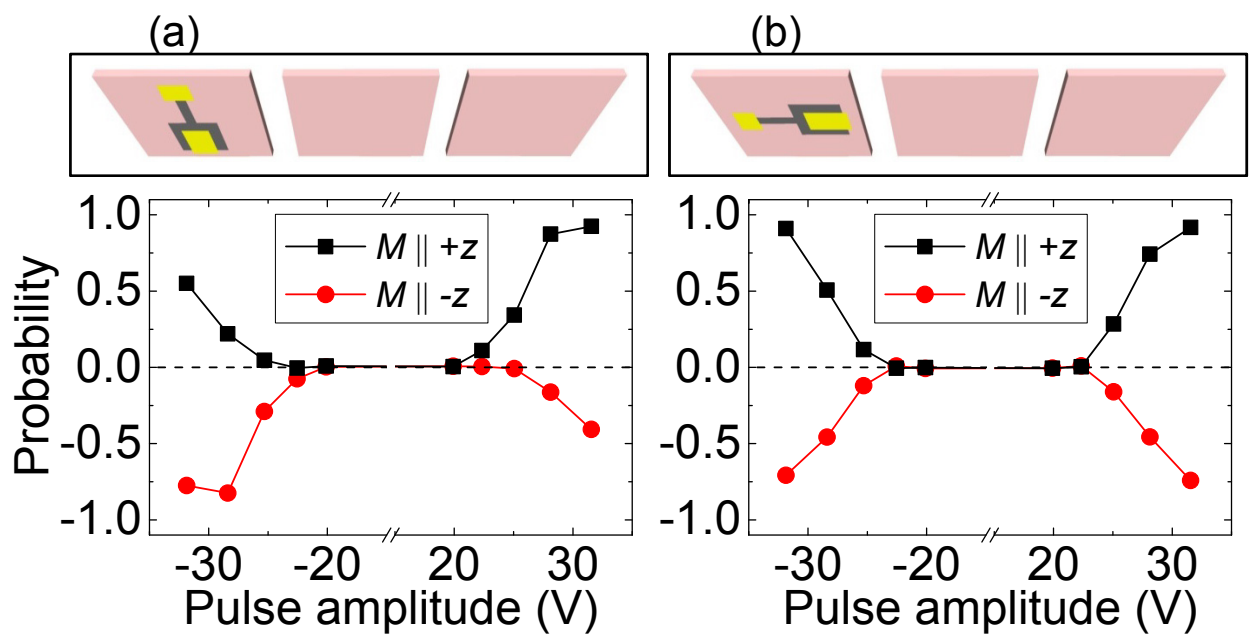

Fig. 6

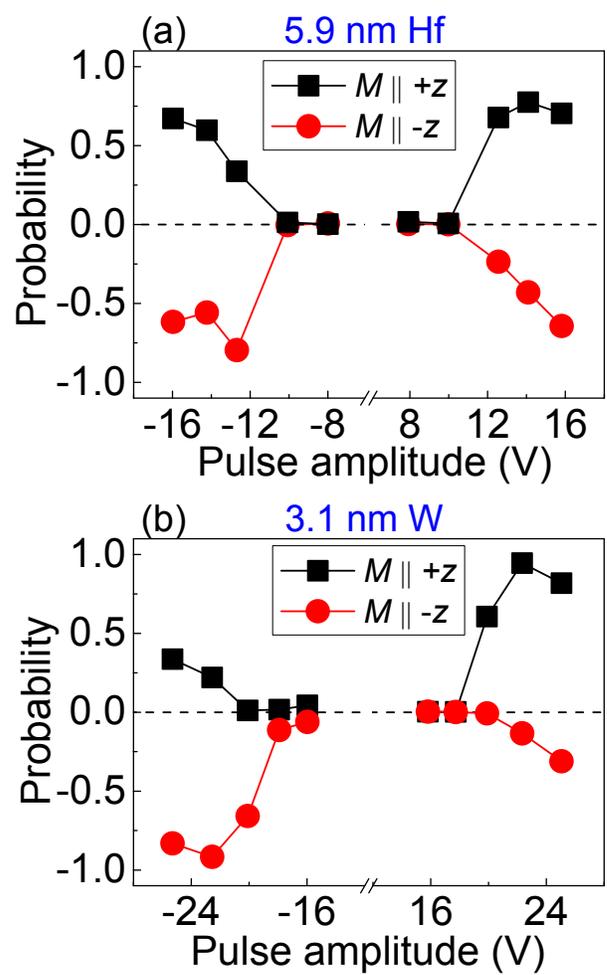

Fig. 7

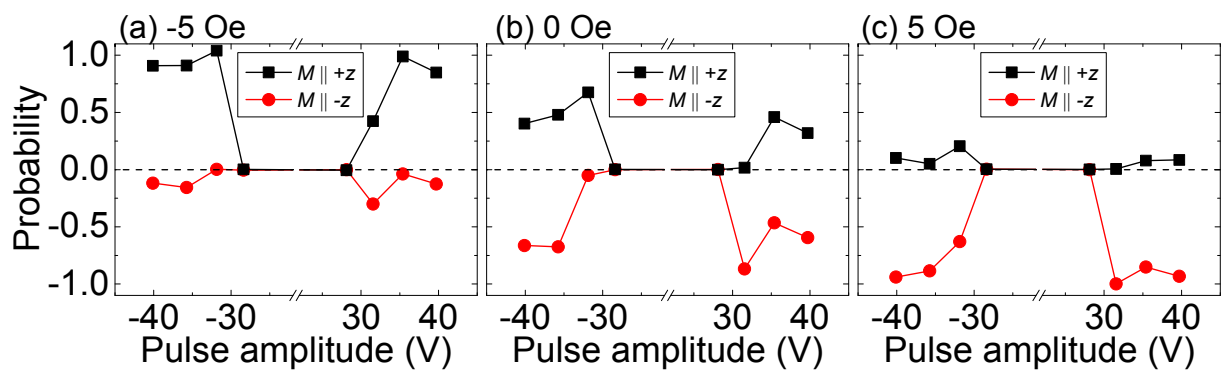

Fig. 8

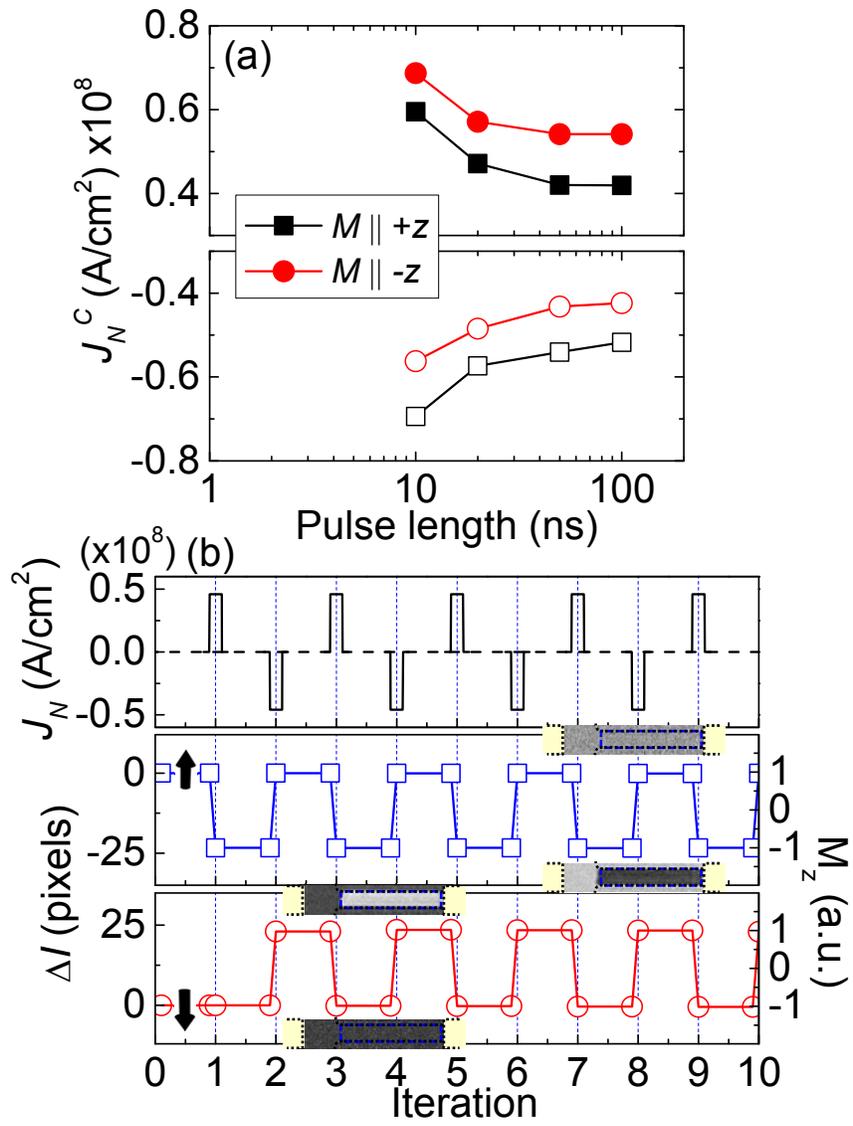

Fig. 9

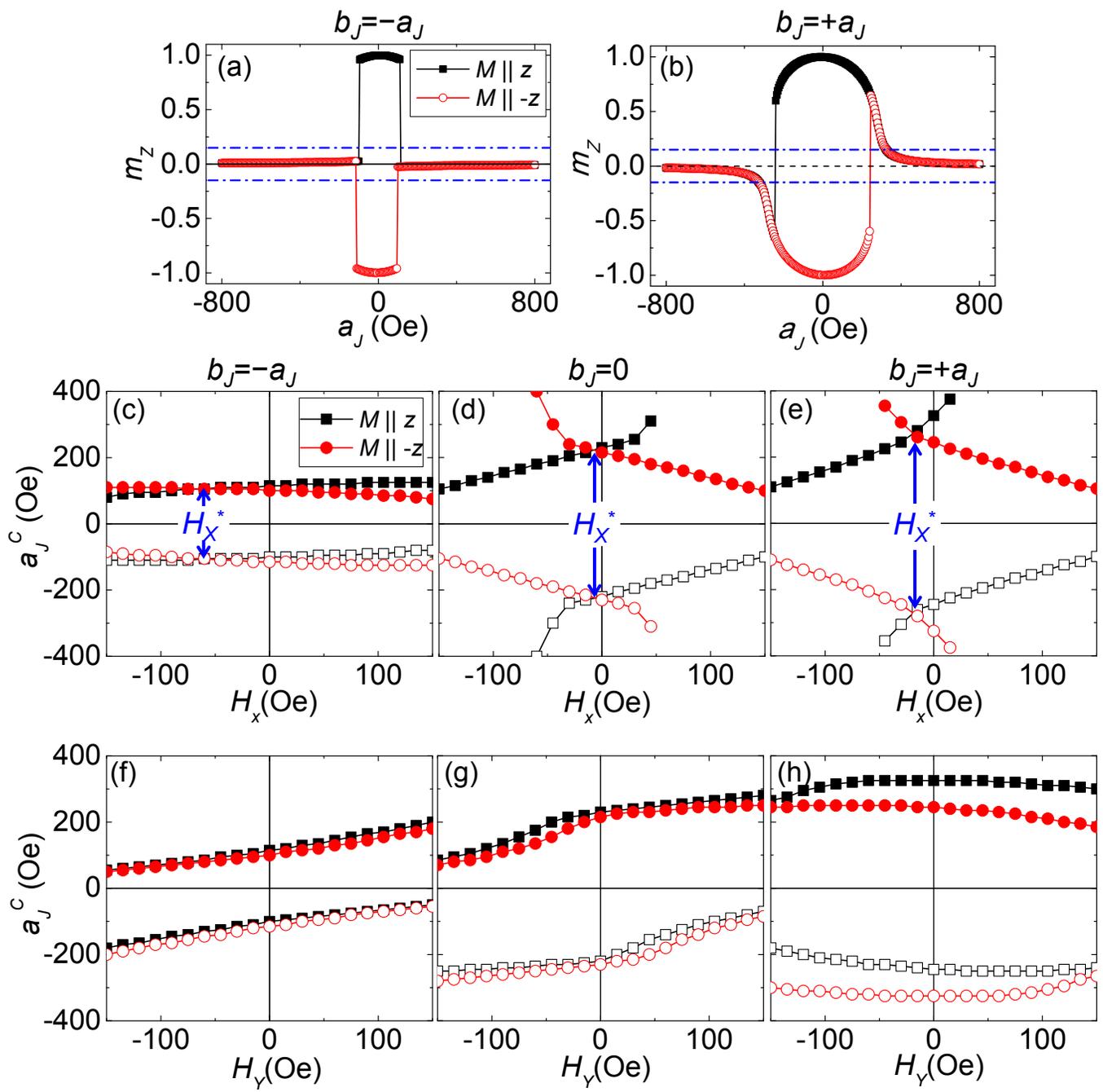

Fig. 10

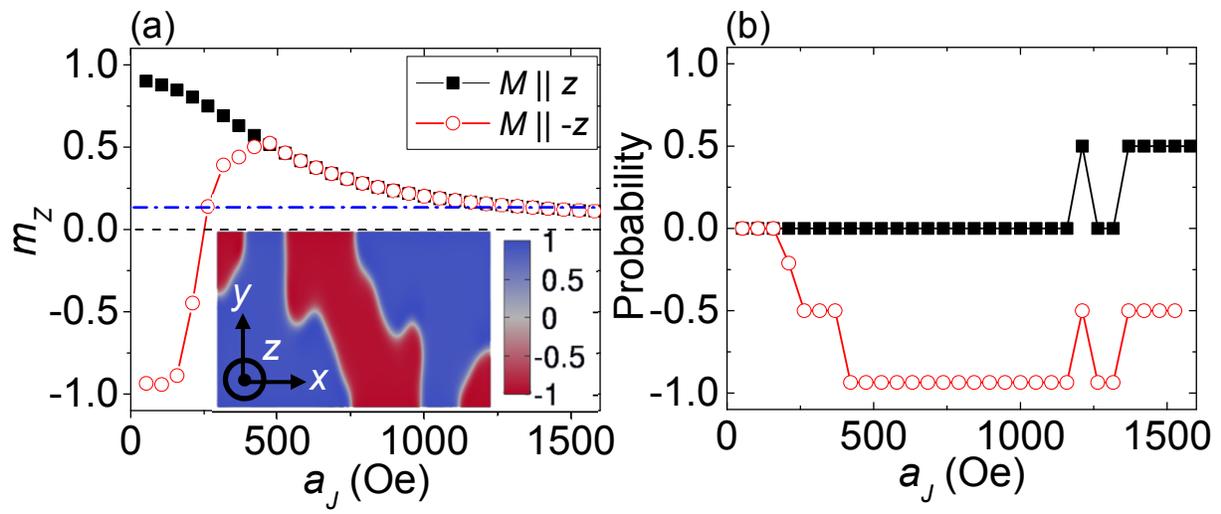

Fig. 11